\RequirePackage{fix-cm}
\documentclass{svjour3}                     % onecolumn (standard format)
\smartqed  % flush right qed marks, e.g. at end of proof
\usepackage{graphicx}
\usepackage[strings]{underscore}
\usepackage{comment}
\usepackage{pifont}
\usepackage[
    colorlinks=true,
    citecolor=blue,
    linkcolor=blue, 
    urlcolor=blue, 	
    pdfstartview=FitH
]{hyperref}

\begin{document}

\title{Graph embeddings for Abusive Language Detection%\thanks{Grants or other notes
}


\author{Noé Cécillon* \and Vincent Labatut \and Richard Dufour \and Georges Linarès}


\institute{Noé Cécillon \at Laboratoire Informatique d’Avignon – LIA EA 4128, Avignon Université, France
              \email{noe.cecillon@univ-avignon.fr}
           \and
           Vincent Labatut \at Laboratoire Informatique d’Avignon – LIA EA 4128, Avignon Université, France
              \email{vincent.labatut@univ-avignon.fr}
           \and
           Richard Dufour \at Laboratoire Informatique d’Avignon – LIA EA 4128, Avignon Université, France
              \email{richard.dufour@univ-avignon.fr}
           \and
           Georges Linarès \at Laboratoire Informatique d’Avignon – LIA EA 4128, Avignon Université, France
              \email{georges.linares@univ-avignon.fr}
}

\date{Received: date / Accepted: date}

\maketitle

\begin{abstract}
Abusive behaviors are common on online social networks. The increasing frequency of anti-social behaviors forces the hosts of online platforms to find new solutions to address this problem. Automating the moderation process has thus received a lot of interest in the past few years. Various methods have been proposed, most based on the exchanged content, and one relying on the structure and dynamics of the conversation. It has the advantage of being language-independent, however it leverages a hand-crafted set of topological measures which are computationally expensive and not necessarily suitable to all situations.
In the present paper, we propose to use recent graph embedding approaches to automatically learn representations of conversational graphs depicting message exchanges. We compare two categories: node vs. whole-graph embeddings. We experiment with a total of 8 approaches and apply them to a dataset of online messages. We also study more precisely which aspects of the graph structure are leveraged by each approach. Our study shows that the representation produced by certain embeddings captures the information conveyed by specific topological measures, but misses out other aspects.
\end{abstract}

%%%%%%%%%%%%%%%%%%%%%%%%%%%%%%%%%%%%%%%%%%%
\section{Introduction}
\label{sec:Intro}
In recent years, online social media have allowed people to meet and discuss world-wide. These popular platforms attract more and more users, and are confronted with an increasing number of abusive behaviors. This phenomenon started to draw attention from governments, requesting companies to perform moderation on their social media platforms. Depending on the size of the communities to be administered, this could be an expensive process since moderation is currently mainly done by human operators. Moreover, this task is difficult, especially because the definition of what constitutes an abuse is ambiguous and can vary depending on the context (\textit{e.g.} media platform, community, and/or country). In order to automate the detection of abusive content in such social media, researchers have proposed methods primarily based on Natural Language Processing (NLP) approaches. These works rely on the textual content of the exchanged messages to detect specific types of abuse, such as offensive language~\cite{Chen2012,Nobata2016,Xiang2012,OkkyIbrohim2018}, hate speech~\cite{Djuric2015,Salminen2018,Warner2012,Badjatiya2017}, racism~\cite{Waseem2016} and cyber-bullying~\cite{Dinakar2011}. However, a major limitation of such NLP-based approaches is their sensitivity to intentional text obfuscation done by malicious users to fool automatic systems. For instance, \texttt{sh1t} and \texttt{f*ck} are easily understandable by humans but difficult to detect by an algorithm if the training corpus does not reflect such situations. Therefore, NLP statistical systems cannot reflect all the forms of language that these abuses can take, due to the wide variety of language registers (which can range from colloquial to sustained), the language proficiency of the contributors, and even the particular vocabulary inherent to the concerned community of users.

To address this language limitation and dependency, authors have proposed to incorporate behavioral information about users and the structure of conversations~\cite{Balci2015,Chatzakou2017,Dadvar2013,Mishra2018b,Yin2009} as a way to improve the efficiency of language-based approaches. In a previous work~\cite{Papegnies2019}, we proposed to leverage conversational graph-based features to detect abusive messages in chat logs extracted from an online game. Such conversational graphs model interactions between users (\textit{i.e.} who is arguing with whom?), completely ignoring the language content of the messages. We characterized the structure of these graphs by computing a large set of manually selected topological measures, and used them as features to train a classifier into detecting abusive messages. As we did not know in advance which topological measures are the most discriminative for this task, we had to consider a very large set, and performed feature selection in order to identify the most relevant ones. This constitutes an important limitation of this method, and more generally of such feature engineering approaches. One important drawback is the important run-time caused by the large number of measures to compute. Furthermore, since the set of measures has to be manually constituted by humans, it could be non-exhaustive, missing relevant features for the task at hand, or on the contrary include a lot of redundant information. It is even possible that no measure defined in the literature captures the relevant information to perform the task at hand.

Graph embedding methods automate this graph representation process. They allow representing graphs as low-dimensional vectors while preserving at least a part of their topological properties. On the one hand, these representations are automatically learned, so they do not require to perform any feature selection, and they are much more time-efficient than the approaches described above. On the other hand, unlike standard topological measures, the obtained representations are not directly interpretable in terms of graph structure. It is therefore not straightforward to understand exactly which information is captured by the embedding, and is possibly relevant to the application. Moreover, different embedding methods are assumed by construction to capture different aspects of the graph structure, but it is difficult to compare them directly, for the same reason. One way to assess the appropriateness of an embedding method for a task, and to compare several embedding methods through on it, is to do so empirically. 

In~\cite{Goyal2018}, Goyal \& Ferrara propose such an experimental work. They compare five methods on tasks such as unobserved link prediction and node classification. They conduct their experiments on various types of networks (\textit{e.g.} social relationships, user network, collaboration network). Nonetheless, only node-level methods are tested and, as stated in~\cite{Cai2018}, performances of graph embedding methods are very task-dependent. Therefore, most effective methods in~\cite{Goyal2018} might not be as appropriate on the task we focus on in this work. Mishra \textit{et al.}~\cite{Mishra2018b} propose to profile authors in order to enhance the detection of abusive content online. They construct a community graph representing all authors and their connections and use a node embedding method to obtain a vector representation of each user called \textit{user profile}. This method shows promising results when combined with standard abuse detection methods relying exclusively on the textual content. It is however limited to the use of a single embedding method.

In this work, we adopt an approach similar to Goyal \& Ferrara and apply it to our abuse detection task. We leverage the (already mentioned) framework presented in our previous work~\cite{Papegnies2019}, which is able to classify messages depending only on the structure of the conversation surrounding them. Text is not used in the process, only \textit{conversational networks}, which makes it language-independent. On this basis, our first contribution is to study the effectiveness of graph embeddings in the context of online abuse detection. We assess and compare 8 methods designed to operate at different scales of the graph (node and whole-graph), and to preserve different structural properties. Our second contribution is an analysis of our results aiming at better understanding which structural properties of the graph are well captured by the considered embedding methods.

The rest of this article is organized as follows. First, in Section~\ref{sec:relatedwork}, we review the literature related to node and whole-graph embedding methods. Then, we present our task in Section~\ref{sec:methodology}, as well as the baseline that we previously developed and the embedding methods that we use in our experiments. In Section~\ref{sec:experiments}, we describe our dataset, our experimental protocol and settings, the results that we obtain, and we discuss the topological properties preserved by each considered embedding method. Finally, in Section~\ref{sec:conclusion}, we summarize our main findings and present some perspectives.

%%%%%%%%%%%%%%%%%%%%%%%%%%%%%%%%%%%%%%%%%%%%%%%%%%%%%%%%%%%%%%%%%%%%
\section{Related Work}
\label{sec:relatedwork}
Generally speaking, the expression \textit{graph embedding} refers to a family of methods aiming to represent graphs, or parts of graphs, in a low-dimensional space in which at least certain aspects of their structure are preserved~\cite{Cai2018}. By construction, objects which are similar for these aspects have close vector representations in the embedding space~\cite{Yan2007}. In addition to the plain structure, certain methods are able to capture additional information such as node labels or the weight and direction of edges.

One can distinguish four main categories of graph embedding methods, depending on the nature of the considered objects: \textit{node} embedding, \textit{edge} embedding, \textit{subgraph} embedding and \textit{whole-graph} embedding. Each category better fits the needs of different applications and problems. In this work, our task can be formulated as a node and/or graph classification problem, hence in the rest of this paper, we focus exclusively on both of these types of embeddings.

The rest of this section is a review of the main node and whole graph embedding methods. Table~\ref{tab:RecapBib} summarizes these methods, and show their main characteristics.

%%%%%%%%%%%%%%%%%%%%%%%%%%%%%%%%
\subsection{Node Embedding}
Node embedding is the most common form of graph embedding in the literature. Such methods take a graph as input, for instance as an adjacency matrix, and output a vector of fixed dimension for each node in the graph. Following the taxonomy proposed in~\cite{Goyal2018}, we distinguish three categories, depending on the general approach used to perform the transformation: Matrix Factorization, Neural Network and Random Walks. Note that the latter also uses neural networks but introduces a different strategy to sample the graph.
%They are used in a variety of applications including link prediction~\cite{Liang2019}, node classification~\cite{Hou2019,Lu2003} and node clustering~\cite{Shi2000,Ding2001}.

%%%
\paragraph{Matrix Factorization}
There are various ways to represent a graph in a matrix form, such as the adjacency, Laplacian or transition matrices. The pioneering studies on node embedding propose to map nodes into low-dimensional vectors by decomposing such matrices into products of smaller matrices of the desired dimension, a process called \textit{Matrix Factorization} (MF).

The most straightforward approach is to leverage existing dimensionality reduction techniques, originally designed for tabular data, and apply them to a graph matrix. Doing so with the \textit{Locally Linear Embedding} (LLE) method proposed by Roweis \& Saul~\cite{Roweis2000} amounts to considering that every node in the graph is a weighted linear combination of its neighbors. The method first estimates weights that best reconstruct the original characteristics of a node from its neighbors, and then uses these weights to generate vector representations. This method has been used in the literature to perform face recognition~\cite{Yan2007}. Belkin \& Niyogi propose \textit{Laplacian Eigenmaps} (LE)~\cite{Belkin2002}, a method aiming at keeping strongly connected nodes close in the result space. Representations are obtained by computing the Eigenvectors of the graph Laplacian. Typical applications for this method include node classification and link prediction~\cite{Belkin2002,Goyal2018}. A major drawback of these two methods is their important time complexity, making them poorly scalable and impossible to use on very large real-world graphs. 

Ahmed \textit{et al.} propose a method called \textit{Graph Factorization} (GF)~\cite{Ahmed2013} which is much more time efficient and can handle graphs with several hundred million nodes. GF uses stochastic gradient descent to optimize the matrix factorization. To improve its scalability, GF uses some approximation strategies, which can introduce noise in the generated representations. Furthermore, GF focuses on preserving only the first-order proximity, \textit{i.e.} nodes which are directly connected have close representations. Hence, the global graph structure is not necessarily well preserved by this method. Ahmed \textit{et al}. use this method to partition graphs and to predict the volume of e-mail exchanges between pairs of users~\cite{Ahmed2013}.

Ou \textit{et al.} introduce a MF method called \textit{High-Order Proximity preserved Embedding} (HOPE)~\cite{Ou2016}. This similarity matrix is obtained using centrality measures like Rooted PageRank, Katz measure and Adamic-Adar score. HOPE is specifically designed to preserve asymmetric transitivity in directed graphs. To this end, two vector representations are learned for each node, a \textit{source} vector and a \textit{target} vector. Applications of this method includes link prediction, proximity approximation and vertex recommendation~\cite{Liang2019}. However, once again the time complexity of this MF method is high and does not allow the processing of very large graphs.

Li \textit{et al.} present \textit{BoostNE}~\cite{Li2019}. This multi-level graph embedding framework learns multiple graph representations at different granularity levels. Inspired from boosting, it is built on the assumption that multiple weak embedding can lead to a stronger and more effective one. It applies an iterative process to a closed form node connectivity matrix. This process successively factorizes the residual obtained from the previous factorization, to generate increasingly finer representations. The sequence of representations produced is then assembled to create the final embedding. Li \textit{et al}. apply their method to a multi-label node classification task.

%%%
\paragraph{Neural Networks}
Neural approaches have been successfully adapted to many fields including graph embedding. Wang \textit{et al.} propose the \textit{Structural Deep Network Embedding} (SDNE) framework~\cite{Wang2016}. This method learns representations based on first and second order proximities in the graph. These two properties are jointly optimized using a deep autoencoder and a variation of Laplacian Eigenmaps, applying a penalty when similar nodes are mapped far from each other in the embedding space. This allows a good representation of both the local and global structure of the graph. This method has been used on tasks similar to the embedding method LE, \textit{i.e.} node classification and link prediction~\cite{Goyal2018,Wang2016}.

Kipf \& Welling develop a method called \textit{Graph Convolutional Networks} (GCN)~\cite{Kipf2017}. It uses an iterative process wherein each iteration captures local neighborhood, and their repetition allows capturing the global neighborhood of nodes. At each iteration, the process aggregates the representations of neighboring nodes and uses a function of the obtained representation and the embedding at previous iteration to generate the new representation. Kipf \& Welling leverage their method to perform document and entity classification.

\textit{Generative Adversarial Networks} (GAN) have also been adapted to node embedding. Wang \textit{et al.}~\cite{Wang2018} propose \textit{GraphGAN}, which works through two models. First, a generator $G(v|v_c)$ tries to approximate the true connectivity between nodes $v$ and $v_c$ and selects the most likely connected nodes to $v_c$. Second, a discriminator $D(v,v_c)$ computes the probability of an edge to exist between $v$ and $v_c$. The generator tries to fit the distribution of nodes as much as possible to generate the most indistinguishable fake pairs of connected nodes. The discriminator tries to distinguish between ground truth and the fake pairs created by the generator. This method is however only able to capture the local structure. Wang \textit{et al}. apply \textit{GraphGAN} to node classification, link prediction and movie recommendation tasks.

%%%
\paragraph{Random Walks}
Random-walks have first been adopted by graph embedding approaches trying to mimic word-embedding methods such as \textit{word2vec}~\cite{Mikolov2013}. They allow representing the graph structure under a sequential form, analogous to sentences in a text. They are used to sample the graph, and can been seen as a proxi allowing to obtain a partial representation of its structure. They also have the advantage of being able to deal with graphs too large to be explored in their entirety. Given a starting node, random-walk-based methods generate node sequences by selecting a neighbor and repeating this procedure until the node sequence reaches a certain length.

Perozzi \textit{et al.} propose \textit{DeepWalk}~\cite{Perozzi2014}. It is among the first node embedding methods based on random-walks. First, DeepWalk samples node sequences using uniform random walks and then applies the standard \textit{SkipGram} model~\cite{Mikolov2013} to generate the representations. This model takes a node as input and aims at predicting its context, \textit{i.e.} the nodes in its neighborhood. With this method, nodes with similar contexts share similar representations. Typical applications of this approach include node classification~\cite{Hou2019,Wang2016} and link prediction~\cite{Ou2016}. However, a limitation is that two nodes can be structurally similar (\textit{i.e.} they play the same role in the graph) but be distant in the graph, hence, not share any common neighbors. Their representations might thus be completely different. 

The \textit{Node2vec}~\cite{Grover2016} method proposed by Grover \& Leskovec was developed following the idea of \textit{DeepWalk}. The main difference is that \textit{Node2vec} uses \textit{biased} random-walks to provide a more flexible notion of a node's neighborhood and better integrate the notion of structural equivalence. It has been used to predict links in a biomedical context~\cite{Liang2019}, and to classify nodes~\cite{Hou2019}. \textit{Node2vec} randomly initializes the node embeddings, which can result in being stuck in a local optima during the computation of embeddings. Chen \textit{et al.} propose an improved weight initialization strategy to avoid such problems in their method \textit{Hierarchical Representation Learning} method (HARP)~\cite{Chen2018}. In \textit{Walklets}~\cite{Perozzi2017}, Perozzi \textit{et al.} introduce a new random walk strategy. Traditional random-walk methods select the next node from the current node's neighbors. Instead, \textit{Walklets} proposes to skip over nodes to obtain sequences of nodes which are not direct neighbors. This strategy allows modeling and preserving higher order relationships between nodes and can be used in multi-label classification problems~\cite{Perozzi2017}.

\begin{table*}
    \caption{List of graph embedding approaches and the additional information they can encode. PyrGE and Graph2vec can additionally handle node attributes and node/edge labels, respectively. Column \textit{Sc.} stands for \textit{Sc}ale (\textit{N}ode vs. \textit{W}hole \textit{G}raph); and \textit{Cat.} for \textit{C}ategory (\textit{M}atrix \textit{F}actorization, \textit{N}eural \textit{N}etworks and \textit{R}andom \textit{W}alks). Columns \textit{W.} and \textit{D.} indicate whether the method supports weighted and directed links, respectively.}
    \label{tab:RecapBib}
    \begin{tabular}{lllllll}
        \hline\noalign{\smallskip}
        \textbf{Sc.} & \textbf{Cat.} & \textbf{Method} & \textbf{Ref.} & \textbf{W.} & \textbf{D.} & \textbf{Application} \\
        \noalign{\smallskip}\hline\noalign{\smallskip}
        N & MF & LLE & \cite{Roweis2000} & -- & -- & Image processing \\
        & & LE & \cite{Belkin2002} & \ding{51} & -- & Node classification, link prediction \\
        & & GF & \cite{Ahmed2013} & \ding{51} & -- & Graph partitioning \\
        & & HOPE & \cite{Ou2016} & \ding{51} & \ding{51} & Node recommendation, proximity approximation \\
        & & BoostNE & \cite{Li2019} & \ding{51} & -- & Node classification \\
        \noalign{\smallskip}\cline{2-7}\noalign{\smallskip}
        & NN & SDNE & \cite{Wang2016} & \ding{51} & -- & Node classification, link prediction \\
        & & GCN & \cite{Kipf2017} & \ding{51} & -- & Document/entity classification\\
        & & GraphGAN & \cite{Wang2018} & -- & -- & Movie recommendation \\
        \noalign{\smallskip}\cline{2-7}\noalign{\smallskip}
        & RW & DeepWalk & \cite{Perozzi2014} & -- & -- & Node classification, link prediction \\
        & & Node2vec & \cite{Grover2016} & \ding{51} & \ding{51} & Node classification, link prediction \\
        & & HARP & \cite{Chen2018} & -- & -- & Node classification \\
        & & Walklets & \cite{Perozzi2017} & -- & -- & Node classification \\
        \noalign{\smallskip}\hline\noalign{\smallskip}
        WG & MF & SF & \cite{deLara2018} & -- & -- & Molecular property prediction \\
        & & PyrGE & \cite{Mousavi2017} & -- & -- & Graph classification \\%Numerical and categorical node attributes & \\
        & & FGSD & \cite{Verma2017} & \ding{51} & -- & Graph classification \\
        & & NetLSD & \cite{Tsitsulin2018} & -- & -- & Graph classification, Community detection\\
        \noalign{\smallskip}\cline{2-7}\noalign{\smallskip}
        & NN & Graph2vec & \cite{Narayanan2017} & -- & -- & Graph visualization, similarity ranking \\
        \noalign{\smallskip}\hline
    \end{tabular}
\end{table*}

%%%%%%%%%%%%%%%%%%%%%%%%%%%%%%%%
\subsection{Whole-Graph Embedding}
As mentioned before, node embedding methods are the most widespread in the literature. But some tasks require information at a higher granularity, in which case one would turn to whole-graph embedding. These methods allow to represent a whole graph as a single vector of fixed length. They take a collection of graphs, and output a representation for each of them.

de Lara \& Pineau~\cite{deLara2018} propose a \textit{S}imple and \textit{F}ast (SF) algorithm based on the spectral factorization of the graph Laplacian. It computes the $k$ smallest positive Eigenvalues of normalized Laplacian of the input graph in ascending order, to form the representation of the whole-graph. de Lara \& Pineau use their approach to predict the properties of chemical compounds~\cite{deLara2018}.

Mousavi \textit{et al.}~\cite{Mousavi2017} introduce a whole-graph embedding hierarchical framework called \textit{Pyramidal Graph Embedding} (PyrGE), based on some ideas originating from image processing algorithms. Important global information from images can be extracted by recursively analyzing local information. In the context of graphs, this means that the overall graph structure can be modeled by analyzing substructures at different scales. To this end, a graph pyramid is formed with subgraphs of different scales. Every graph is embedded into vector representations which are all concatenated to form the global graph embedding. The representations are obtained by factorizing an affinity matrix. PyrGE is especially designed for large graphs since they potentially contain more different scales. Mousavi \textit{et al.} used it for graph classification tasks.

Verma \& Zhang propose a \textit{Family of Graph Spectral Distances} (FGSD)~\cite{Verma2017} to represent a whole-graph. This method is built on the assumption that the graph atomic structure is encoded in the multiset of all node pairwise distances. It computes the Moore-Penrose Pseudoinverse spectrum of the graph Laplacian. Vector representation of the whole graph is constructed from the histogram of this spectrum. Typical tasks include graph classification in various fields such as bioinformatics and social networks~\cite{Verma2017,Tsitsulin2018}.

Tsitsulin \textit{et al.} introduce \textit{NetLSD}~\cite{Tsitsulin2018}, a permutation- and size-invariant, scale-adaptive embedding method. Like aforementioned node embedding method LE~\cite{Belkin2002}, NetLSD operates on the Laplacian matrix of the graph. It relies on a physical analogy consisting in simulating a heat diffusion process on the graph to preserve its structure. The method processes the amount of heat transferred between nodes at different times scales. These heat traces at different time scales are then used to compute the heat trace signature of the graph, \textit{i.e.} the vector representation of the graph. Tsitsulin \textit{et al.} use \textit{NetLSD} for graph classification and for community detection. 

Narayanan \textit{et al.} design Graph2vec~\cite{Narayanan2017}, which can be viewed as an adaptation of DeepWalk~\cite{Perozzi2014} and Node2vec~\cite{Grover2016} to the whole-graph embedding paradigm. Indeed, these two approaches generate random walks to approximate the context in which nodes appear and fetch them to a \textit{SkipGram} model. Graph2vec also uses a \textit{SkipGram} model, but it operates on rooted subgraphs since the method is aimed at representing whole graphs and not nodes. Hence, similarly to nodes with similar neighborhoods sharing close representations in DeepWalk and Node2vec, graphs containing the same rooted subgraphs share similar representations in Graph2vec. A \textit{SkipGram} model is then trained on these subgraphs and generates the whole graph representations. \textit{Graph2vec} has been used to perform graph classification, graph visualization and similarity ranking~\cite{Bai2019}. It is able to capture information about node labels additionally to the graph structure.

%%%%%%%%%%%%%%%%%%%%%%%%%%%%%%%%%%%%%%%%%%%%%%%%%%%%%%%%%%%%%%%%%%%%%%%%%
\section{Methods}
\label{sec:methodology}
In this work, we focus on a task consisting in detecting abusive messages in chat logs. This can be formulated as a classification problem consisting in deciding if a message is abusive or not. In order to turn a chat log into a graph, we rely on a conversational graph extraction method that we previously introduced in~\cite{Papegnies2019}, and that we briefly present in Section~\ref{subsec:GraphExtraction}. The principle here is that classifying the messages amounts to classifying the graphs that represent them.

This setup allows us to experiment with various node and whole-graph embedding methods, which we present in Section~\ref{subsec:embeddings}. For comparison, we use as a baseline a set of features that we manually crafted in a previous work~\cite{Papegnies2019}. These are constituted of a large set of topological measures, that we selected to get the most exhaustive representation of the graph that we could, as explained in Section~\ref{subsec:baseline}. We view the embedding methods as a way to \textit{automate} the elaboration of this representation of conversational graphs, which was otherwise designed \textit{manually} in~\cite{Papegnies2019} through feature selection.

\begin{figure*}[htb]
    \center
    \includegraphics[width=1\textwidth]{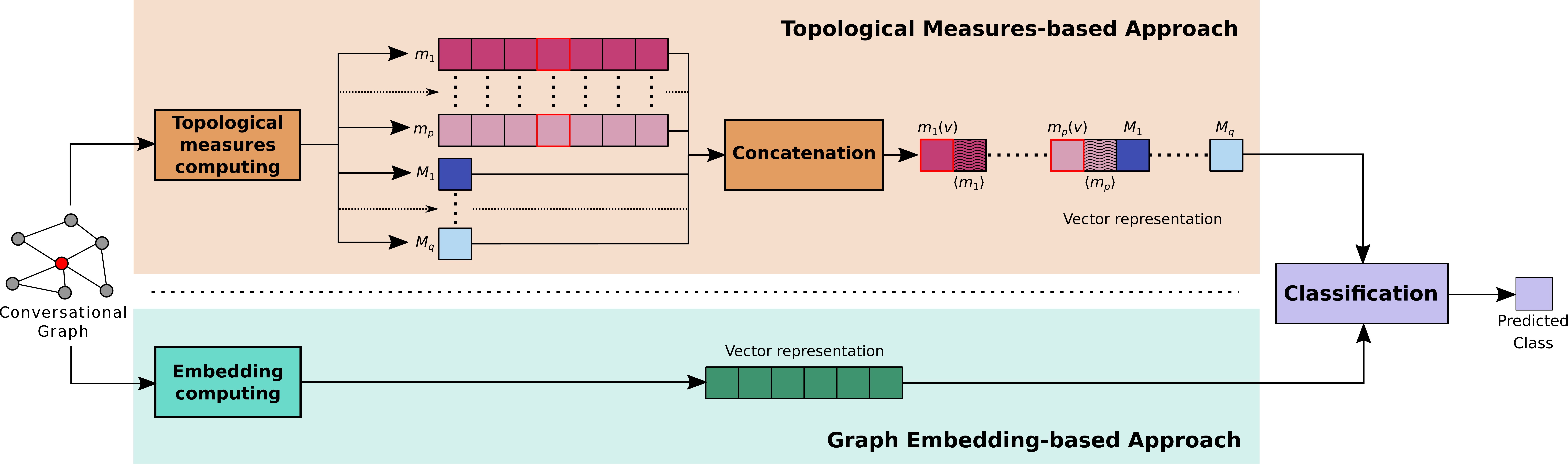}
    \caption{Overview of our experimental framework. The top part corresponds to the approach adopted in our baseline, whereas the bottom part describes the method used with graph embeddings. Figure available at \href{https://doi.org/10.6084/m9.figshare.7442273}{10.6084/m9.figshare.7442273} under CC-BY license.}
    \label{fig:embedding}
\end{figure*}

Figure~\ref{fig:embedding} gives an overview of our experimental framework, highlighting the differences between the approaches based on the topological measures (top) and the embedding methods (bottom). The baseline features are computed separately for the input graph, before being concatenated to form the global representation of the graph, and this single vector is finally fetched to the classifier. By comparison, the graph embedding method directly produces a single vector representation of fixed length (6 in this example) which is then sent to the classifier in a straightforward way.

%%%%%%%%%%%%%%%%%%%%%%%%%%%%%%%%%
\subsection{Graph Extraction}
\label{subsec:GraphExtraction}
Intuitively, the content exchanged in an online conversation could be assumed to be the most relevant information to detect important events, such as the occurrence of abuses. However, we have showed in a previous work~\cite{Papegnies2019} that the dynamics of the conversation, \textit{i.e.} the way the interactions between its participants unfold, is also critical, and can even lead to better classification results. This information can be leveraged by modeling the exchanges between participants through a so-called \textit{conversational graph}. In this work, we use the same method to extract graphs from conversations. We explain the most essential points of this process in the rest of this section, but the interested reader will find a more detailed description in~\cite{Papegnies2019}.

Our method is designed to process a stream of messages posted in a given chatroom. It extracts a graph describing the conversational context of a message of interest, called \textit{targeted message}. In the context of a classification task, this message corresponds to the message that one wants to classify. We define a so-called \textit{context period} centered around this targeted message, and containing the $k$ messages published right before and right after it (where $k$ is a predefined constant). The graph corresponds to the temporal integration of this events occurring during this period. Each one of its nodes represents a participant of the conversation, which was active at least once during the considered context period. The graph links are directed and weighted. They model the interactions between these participants over the period: their directions reflect the communication flow, and their weights represent the overall intensity of the exchanges. The iterative process used to estimate the link directions and weights is detailed in~\cite{Papegnies2019}, as well as the parameters that can be used to control this process.

\begin{figure}
    \center
    %0.75
    \includegraphics[width=0.75\textwidth]{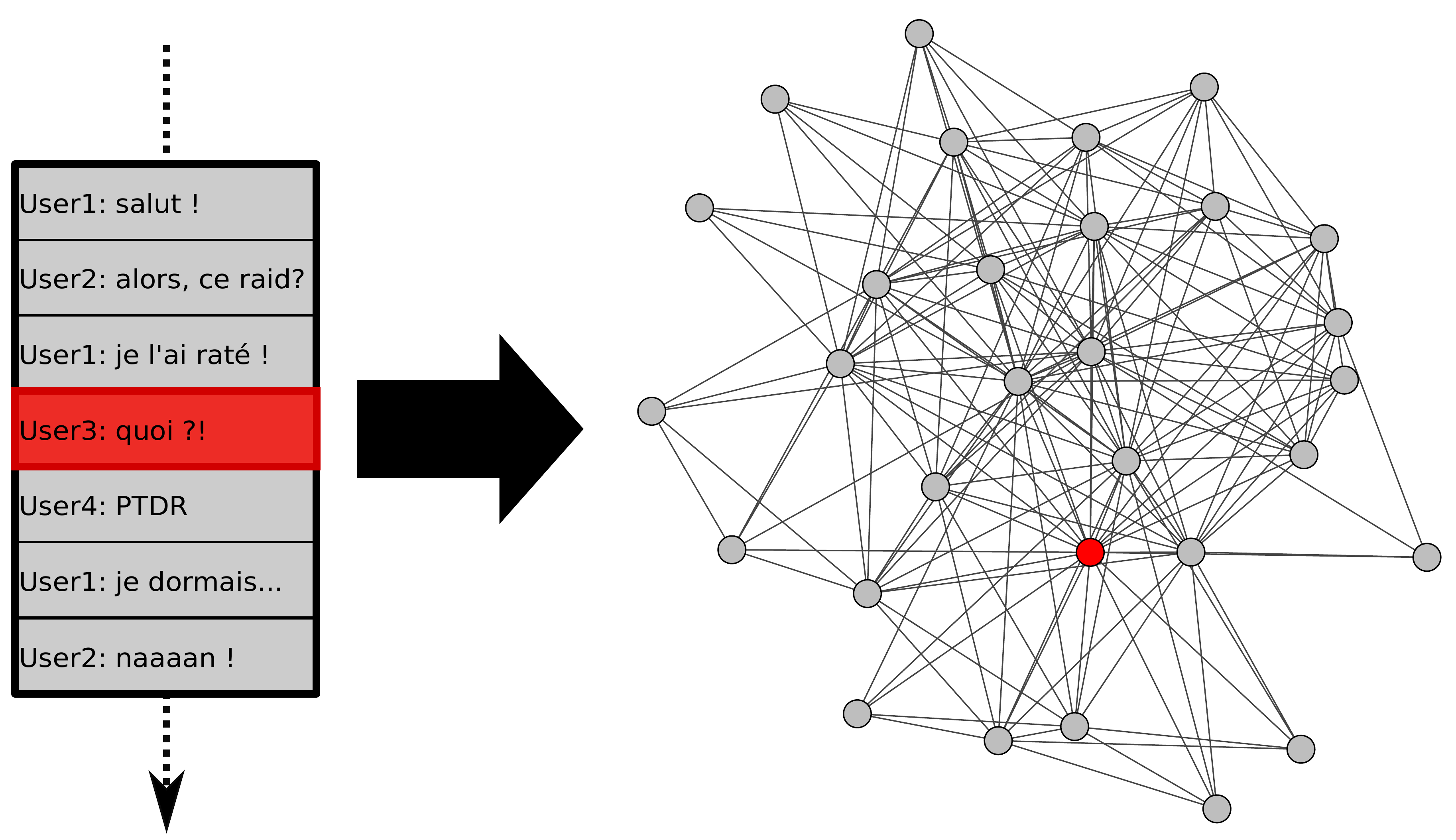}
    \caption{Representation of our method to build graphs from conversations. The left part is an extract of the considered conversation, which takes the form of a sequence of chat messages. The red message corresponds to the \textit{targeted message}, \textit{i.e.} the message we ultimately want to classify. The right part is the corresponding conversational graph, with the author of the targeted message in red. For readability reasons, weights and directions have been omitted in the graph. Figure available at \href{https://doi.org/10.6084/m9.figshare.7442273}{10.6084/m9.figshare.7442273} under CC-BY license.}
    \label{fig:ConvGraphs}
\end{figure}

Figure~\ref{fig:ConvGraphs} illustrates our graph extraction method. The left part is the conversation (stream of messages), with the targeted represented in red. The graph representing the state of the conversation around this message is shown on the right. Its red node models the author of the targeted message.

%%%%%%%%%%%%%%%%%%%%%%%%%%%%%%%%
\subsection{Baseline}
\label{subsec:baseline}
Our baseline relies on our previous work presented in~\cite{Papegnies2019,Cecillon2019}. In~\cite{Papegnies2019} we only focus on graph-based features, but in~\cite{Cecillon2019} we also leverage textual content to perform the classification. We propose 3 strategies to combine both types of features: 1) \textit{Early fusion} relies on a global feature set, containing both text- and graph-based features; 2) \textit{Late fusion} uses two separate classifiers dedicated to text- and graph-based features, respectively, and fetches their outputs to a third classifier; and 3) \textit{Hybrid fusion} combines both previous strategies.

However, our goal in the current work is to study the behavior of \textit{graph embedding} methods on this task. Therefore, we \textit{only} focus on the \textit{interactions} between the participants of the conversation, as modeled by the graphs whose extraction process was just described in Section~\ref{subsec:GraphExtraction}. It is worth stressing that completely ignoring the textual content exchanged by the participants of the conversation makes our method language-independent, and obfuscation-resistant.

We select a set of standard topological measures to describe the graph in a number of distinct ways, in terms of scale and scope. The \textit{scale} depends on the nature of the characterized object: node or graph. Some of the measures characterize the graph as a whole (\textit{i.g.} diameter, density), whereas other focus on individual nodes (\textit{i.g.} degree, closeness). The \textit{scope} corresponds to the nature of the information used to characterize the object: micro, meso, or macroscopic. Some of the selected measures leverage only local information (\textit{i.g.} transitivity, reciprocity), other consider the full graph (\textit{i.g.} betweenness, eccentricity) or intermediate substructures (\textit{i.g.} modularity, participation coefficient).

The graph scale measures, denoted $M_1,...,M_q$ in Figure~\ref{fig:embedding}, are directly used as classification features. The node measures, denoted $m_1,...,m_p$ in the same figure, are computed for all nodes, and used to produce two different types of features. The first corresponds to the value obtained for the node modeling the author of the targeted message (\textit{i.e.} red node in Figure~\ref{fig:ConvGraphs}), denoted $m_i(v)$ in Figure~\ref{fig:embedding}. The second is the average of this measure over all nodes in the graph, denoted $\langle m_i \rangle$ in the figure, which is considered as a graph scale representation. In total, the full set is constituted of 459 features, including several variants of certain topological measures. Their detailed list is available in~\cite{Papegnies2019}.

For each annotated message in our corpus, we first extract the corresponding conversational graph, as explained before. We then compute the whole set of topological measures to fully describe each one of these graphs. The graph-scale measures allow to characterize the whole conversation at once, whereas the node-scale measures are used to describe the position of the node corresponding to the author of the targeted message. Finally, all of these values are used as input features fetched to an SVM classifier. In addition, we perform a feature ablation study to identify the most discriminative topological measures for the task at hand, which we call \textit{Top Features}. It turns out that 9 top features are enough to reach a performance 97\% as good as the performance obtained with the \textit{whole} feature set on the test set.

%%%%%%%%%%%%%%%%%%%%%%%%%%%%%%%%
\subsection{Embedding Methods}
\label{subsec:embeddings}
In this subsection, we describe the graph embedding methods that we use in our experiments. We found in our previous work~\cite{Papegnies2019} that topological measures describing the graph at different scales and scopes can convey complementary information, allowing to improve the performance on the classification task. This is the reason why we decided to include both whole-graph and node embedding methods in this study. We selected methods that use different strategies, and focus on preserving various aspects of the graph, in order to include as much diversity as possible. All implementations are from the \textit{Karate club toolkit}~\cite{Rozemberczki2020} except Node2vec, which was developed by E. Cohen\footnote{https://github.com/eliorc/node2vec}. In our description, the names of the parameters correspond to those used in these toolboxes.

%%%%%%%%%%%%%%
\subsubsection{Whole-Graph Embedding}
As explained in Section~\ref{subsec:GraphExtraction}, we extract a conversational graph for each targeted message, based on its context period. Using a description of the whole graph amounts at considering the entire conversation at once when performing the classification. We found in our previous study~\cite{Papegnies2019} that certain graph-scale topological measures such as the \textit{Authority score} and \textit{Reciprocity} are particularly discriminative for the task at hand. In this experiment, we consider whole-graph embedding methods as the embedding analog of graph-scale measures.

%%%
\paragraph{Spectral Features~\cite{deLara2018} (SF)}
This method was developed to perform a classification task over a corpus of unweighted undirected graphs. Moreover, it assumes each graph is connected. Its first step is quite standard and consists in computing the spectrum of the normalized graph Laplacian, keeping only the $k$ smallest positive Eigenvalues. The very smallest of these values is ignored though, as it corresponds to the number of components, i.e. 1 according to the above assumption. 

These Eigenvalues, in ascending order, form the representation of the graph. If the graph contains less than $k$ nodes, (resulting in less than $k$ Eigenvalues), the vector is right-padded with zeros. Parameter $k$, called \texttt{dimensions} in the implementation, therefore directly controls the size of the representation.

%%%
\paragraph{Family of Graph Spectral Distances~\cite{Verma2017} (FGSD)} 
This method was also proposed to perform classification task over a corpus of undirected graphs, but now these are weighted. Verma \& Zhang designed their representation in order to characterize a graph in terms of certain of its constituting subgraphs, and so that it has the property of being invariant under graph isomorphism. It relies on the assumption that the characteristics of the graph are encoded in the set of all its node pairwise distances. They propose a family of graph spectral distances (FGSD) based on the spectrum of the graph Laplacian, which is able to encode both local and global structure properties. They select the most appropriate distance in this family, in order to fulfill their objective of isomorphism-invariance, and to obtain a sparse representation. 

This results in a representation whose length depends on the graph order. To get a fixed-length vector suitable for classification, they discretize the distribution of the obtained node pairwise distances through a histogram. Parameter-wise, the user controls the way this histogram is computed. It is necessary to provide the range covered by the histogram (\texttt{hist\_range}) and its number of bins (\texttt{hist\_bins}).

%%%
\paragraph{Graph2vec~\cite{Narayanan2017} (G2V)}
This method was not designed for a specific task, but was evaluated on graph classification and clustering benchmarks. Unlike the previous methods, graph2vec must be trained on a corpus of graphs, as it relies on unsupervised learning through a neural network. It is designed analogically to document embedding methods proposed in NLP. As these methods leverage the fact that a document is formed of a sequence of words, Narayanan \textit{et al}. consider a graph as the set of subgraphs surrounding each node.

The algorithm takes the set of graphs to represent, and outputs their representations by applying a two-step process. It first identifies the subgraphs surrounding each node and constituting the graph. More precisely, it looks for so-called \textit{rooted} subgraph, \textit{i.e.} node neighborhoods of a certain order. Second, these subgraphs are considered as the vocabulary and fetched to a \textit{doc2vec SkipGram}~\cite{Le2014} model. To reduce the computational cost, the method follows a negative sampling strategy (\textit{i.e.} at each iteration, the model updates the representations of only a fixed number of negative samples). 

This embedding method captures \textit{structural equivalence}, \textit{i.e.} graphs whose nodes tend to possess this form of similarity will be close in the representation space. In addition, it is able to take into account an extra input corresponding to a label associated to each node. Parameter-wise, the user must specify the degree of the rooted-subgraphs (\texttt{wl\_iterations}), while the rest of the parameters controls the SkipGram model: size of the representation (\texttt{dimensions}), down-sampling frequency (\texttt{down\_sampling}), number of epochs (\texttt{epochs}), learning rate (\texttt{learning\_rate}) and minimal count of graph feature occurrences (\texttt{min\_count}).

%%%%%%%%%%%%%%%
\subsubsection{Node Embedding}
In the conversational graph extracted from the context of the targeted message, all nodes are not equal. As mentioned in Section~\ref{subsec:GraphExtraction}, one of them represents the author of the targeted message, which we assume plays a particular role if an abuse is occurring at this moment of the conversation. In~\cite{Papegnies2019} we experimented with a node-based representation of the conversation, consisting in characterizing individually this node of interest (by opposition to the whole graph), through a selection of nodal topological measures such as \textit{Strength} and \textit{Closeness centrality}. The node embedding methods presented in this section can be considered as the embedding analog of these measures in the present study.

%%%
\paragraph{DeepWalk~\cite{Perozzi2014} (DW)}
DeepWalk relies on another analogy between graph and text, allowing to adapt a neural network-based approach originating from NLP. It takes a graph as input and uses a set of random walks of fixed length as a proxy to represent the graph structure. The procedure samples a certain number of uniform random walks starting from each node, which are considered as analog to a set of sentences, whereas the nodes set corresponds to the vocabulary. DeepWalk uses a SkipGram model to update the node representations by predicting their neighborhood (\textit{i.e.} context). The obtained representation captures the modular structure of the graph.

The parameters of this method include the size of the neighborhood that we want to consider (\texttt{window\_size}), the learning rate (\texttt{learning\_rate}), the number of epochs (\texttt{epochs}) and the minimal count of node occurrences for including the node in the model (\texttt{min\_count}). Other parameters correspond to the size of the generated embeddings (\texttt{dimensions}), the number of random-walks starting at each node (\texttt{walk\_number}) and their maximum length (\texttt{walk\_length}). The last two parameters are typical of random walk-based approaches.

%%%
\paragraph{Node2vec~\cite{Grover2016} (N2V)}
Node2vec is designed to preserve the node neighborhood in the space of representation. It follows the main idea of \textit{DeepWalk} but uses \textit{biased} random-walks by introducing weights to the transition probabilities between nodes. The goal with this change is to improve the sampling step and get random walks that better model node neighborhoods. The bias allows controlling the behavior of the random walker, resulting in a trade-off between purely breadth-first (exploring the closer nodes first) and depth-first (favoring increasingly distant nodes) samplings. The former tends to produce representation that preserve structural equivalence, whereas the latter provides a wider view of the neighborhood. 

\textit{Node2vec} has the same parameters as \textit{DeepWalk} plus two extra ones. The return parameter \texttt{p} controls the likelihood of immediately revisiting a node during the walk, and the in-out parameter \texttt{q} controls the balance between the breadth-first and depth-first strategies.

%%%
\paragraph{Walklets~\cite{Perozzi2017} (WL)}
Walklets is an extension of DeepWalk which aims at explicitly modeling multi-scale relationships, i.e. combine distinct views of node relationships at different granularity levels. Walklets introduces a key change in the random walk sampling algorithm, as the walk can now skip some nodes to reach farther parts of the network. This allows reaching distant nodes while keeping walk lengths short and tractable. Implicitly, this amounts to sampling different powers of the adjacency matrix. Like DeepWalk, the random walks are the inputs of a SkipGram model. It creates a representation for each power of the adjacency matrix that is explored (\textit{i.e.} each size of \textit{skip}) and the output representation is the result of their concatenation.

The method has the same set of parameters as \textit{DeepWalk}. However, in Walklets, the window size denotes the power order of the adjacency matrix to use (\textit{i.e.} the size of \textit{skips} in random walks) and thus, the number of distinct representations that the model learns. The dimension corresponds to the size of each representation. The size of the global embedding generated is the product of the values of these two parameters.

%%%
\paragraph{BoostNE~\cite{Li2019} (BNE)}
\textit{BoostNE} also learns multiple graph representations at different granularity levels, but unlike Walklets, it relies on matrix factorization. It applies the principle of gradient boosting to perform successive factorizations of an original target matrix denoted as node connectivity matrix. Each one results in a representation corresponding to an increasingly finer granularity. The final embedding is obtained by concatening these representations.

The parameters of this method are the following. First, the user must specify the number of granularity levels considered (\textit{iterations}), as well as two parameters controlling the non-negative matrix factorization step (\texttt{order} and \texttt{alpha}). Finally, similarly to \textit{Walklets}, parameter \textit{dimensions} corresponds to the size of the representation.

%%%
\paragraph{GraphWave~\cite{Donnat2018} (GW)}
This representation was designed to preserve the structural roles of nodes while being robust to small perturbations in the graph structure. It leverages heat wavelet diffusion patterns to estimate a multidimensional representation. The process mimics a physical process consisting in propagating some energy  through the graph structure, starting from the node of interest. The way this energy is diffused over the graph is assumed to characterize the node and its neighborhood. Formally, it is represented by the distribution of wavelet coefficients, which is sampled to get the proper vector representation of the nodes.

Parameter-wise, a scaling parameter allows controlling the hear kernel (\textit{heat\_coefficient}), which corresponds in terms of graph structure to the radius of the considered node neighborhood. The number of points used to sample the distribution of wavelet coefficients (\textit{sample\_number}) corresponds to the size of the representation. The granularity of the grid used to perform this sampling is controlled by parameter \texttt{step\_size}. Wavelet calculation can be performed exactly of approximately (\texttt{mechanism}), in which case the parameters \texttt{switch} and \texttt{approximation} allow controlling its precision.

%%%%%%%%%%%%%%%%%%%%%%%%%%%%%%%%%%%%%%%%%%%%%%%%%%%%%%%%%%%%%%%%
\section{Experiments}
\label{sec:experiments}
In this section, we  present the data and the experimental protocol necessary to carry out our experiments (Section~\ref{subsec:DataProtocol}), and explain how we fixed the many parameters of the considered graph embedding methods (Section~\ref{subsec:parametersEmbed}). We then seek to compare the performance of various types of graph embedding methods (Section~\ref{subsec:results}). We finally explore the complementarity of such methods with our baseline based on topological measures (Section~\ref{subsec:ablation}).

%%%%%%%%%%%%%%%%%%%%%%%%%%%%%%%%
\subsection{Data and Experimental Protocol}
\label{subsec:DataProtocol}

%%%
\paragraph{Data}
The raw data is a proprietary database containing approximately 4 million messages, written in French and posted on the in-game chat of SpaceOrigin\footnote{\url{https://play.spaceorigin.fr/}}, a Massively Multiplayer Online Role-Playing Game (MMORPG). 

Among all the exchanged messages, 655 have been reported as being abusive by other players, and confirmed as such by at least one human moderator. They constitute the \textit{Abuse} class. Non-abusive messages constitute most of the dataset (more than 99\% of the messages): in this case, it is standard to use only a part of the available data in order to get balanced training and testing classes, and thus prevent the classifier from being biased toward the majority class. We constitute the \textit{Non-abuse} class by randomly sampling the same number of messages from the ones that have not been reported, with the constraint that a message must not appear in the same conversation as an already selected message~\cite{Papegnies2019}. As a result, our dataset is composed of 1,320 independent messages, equally distributed between the \textit{Abuse} and \textit{Non-abuse} classes. Equally distributing the dataset leads to a higher performance and is common in abuse detection literature~\cite{Nobata2016,Salminen2018,Balci2015,Yin2009}.

Additionally, we construct a small development set of 120 messages following the same procedure, meant to be used when estimating the embedding methods parameters. This set is also balanced. We associate each message to its surrounding context (\textit{i.e.} messages posted before and after it), as explained in Section~\ref{subsec:GraphExtraction}.

%%%
\paragraph{Experimental Protocol}
We conduct our experiments on a binary classification task to detect whether a message belongs to the \textit{Abuse} or the \textit{Non-abuse} class. We apply the graph extraction process developed in Section~\ref{subsec:GraphExtraction} to our dataset in order to construct a conversational graph for each one of its messages. It is controlled by certain parameters, for which we use the best value identified in our previous work~\cite{Papegnies2019}, which are: a sliding window of 10 messages, a context period of 850 messages, and link computed through a linear assignment strategy. On average, the extracted graphs contain 46 nodes and 500 edges. We make this dataset of $1,320$ conversational graphs publicly available online\footnote{DOI: \href{https://doi.org/10.6084/m9.figshare.7442273}{\texttt{10.6084/m9.figshare.7442273}}}.

We use the 8 embedding methods presented in Section~\ref{subsec:embeddings} in addition to the baseline approach (topological features) described in Section~\ref{subsec:baseline} to generate vector representations of these graphs. For the whole-graph embedding methods, we use the complete representation, whereas for node embedding methods, we only use the representation of the node representing the author of the targeted message (cf. Section~\ref{subsec:GraphExtraction}).

We input these representations to an SVM to perform the classification. We use the implementation provided by the Sklearn toolkit~\cite{Pedregosa2011}, under the name SVC (C-Support Vector Classification). As an alternative, we also experimented with Sklearn's implementation of the multilayer perceptron (MLP). However, it yields very similar performances compared to the SVM, so we decided to not present them here. We suppose that the size of our dataset lowers the effectiveness of such approaches neural approaches, at least on our task. We conduct our experiments on an Intel Core i3-3250 3.5 GHz CPU. Because of the relatively small size of our dataset, we set-up our experiments using a 10-fold cross-validation. We use 70\% of the data for training and the remaining 30\% for testing. Each fold is balanced in terms of classes. 

The classification performance is expressed in terms of micro $F$-measure, the harmonic mean of the precision and recall. Since our dataset is balanced, micro and macro $F$-measure are equivalent in our experiments. In the rest of the paper, we use $F$-measure to refer to them collectively.

%%%%%%%%%%%%%%%%%%%%%%%%%%%%
\subsection{Parameters of Graph Embedding Approaches}
\label{subsec:parametersEmbed}
In this section, we describe the configuration of the parameters of all graph embedding approaches used in our experiments. We vary the values of parameters on the development set to determine their optimal values. As this is not the main objective of this work, we only focus on the main results here.

\begin{figure}
    \center
    \includegraphics[width=0.75\textwidth]{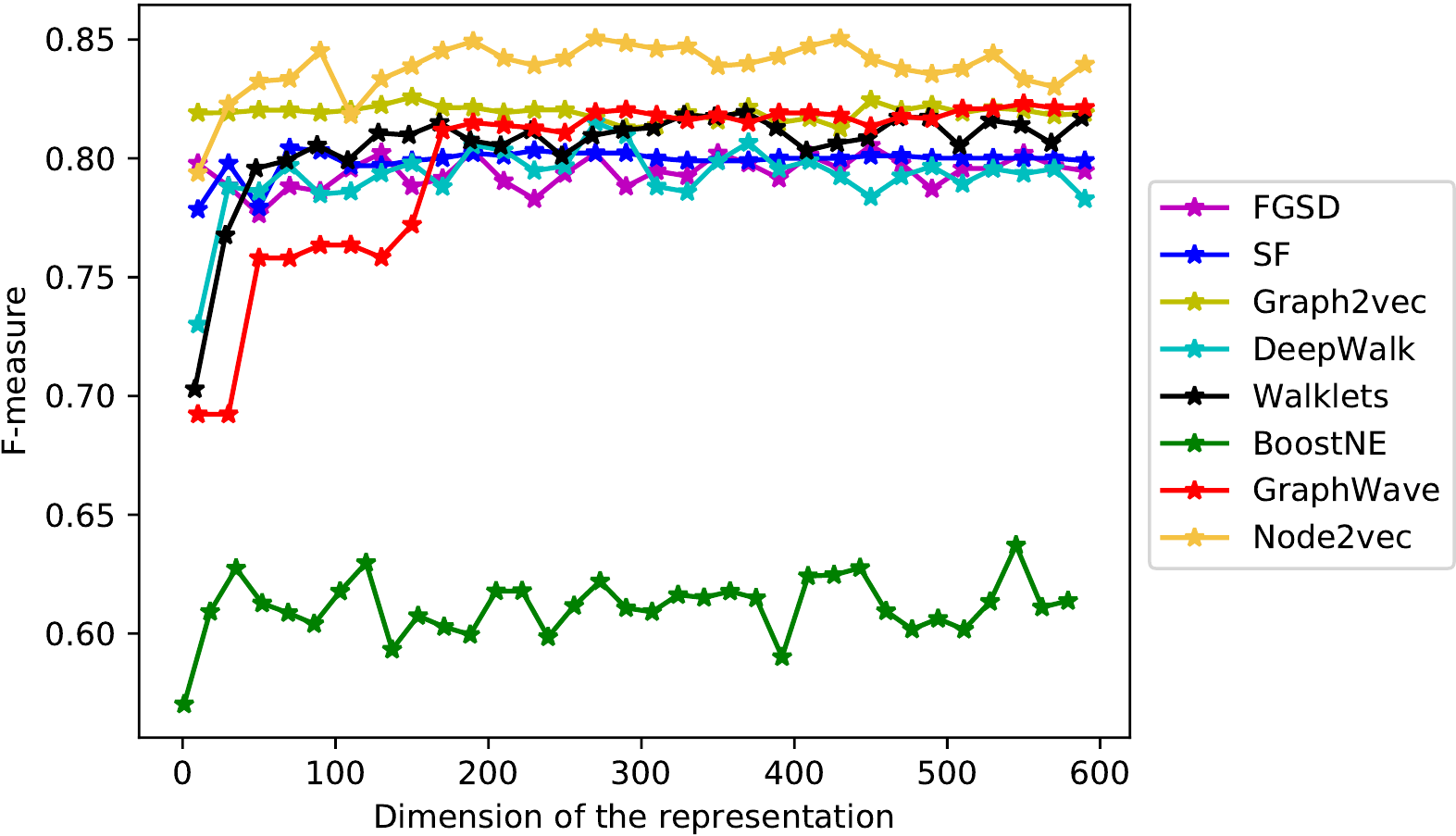}
    \caption{Performance in terms of $F$-measure, as a function of the representation dimension. Figure available at \href{https://doi.org/10.6084/m9.figshare.7442273}{10.6084/m9.figshare.7442273} under CC-BY license.}
    \label{fig:dimensions}
\end{figure}

During our experimentation, we found that most of the parameters have only a limited impact on the performances of embedding methods on our abuse detection task. This includes the dimension of the generated representation, which is the only parameter common to all methods. Figure~\ref{fig:dimensions} shows the performance as a function of this dimension, for all the methods. Performances are computed on the development set. Note that in this figure, we consider the dimension of the output representation (\textit{i.e.} from the extracted embedding representation of graphs) and not the dimension of individual embeddings which are then concatenated in methods such as Walklets and BoostNE. As expected, for most methods, a dimension too small seems to lack discriminative power, as there is not enough information to reliably represent the graph structure. Conversely, as our dataset is composed of relatively small graphs (a few hundred nodes at most), a dimension too large appears does not improve the performance, and just increase the computational cost. Put differently, it seems that we do not need a very large representation of the graph to reach the best performance on this task. This is consistent with our findings from our previous work on the baseline, in which we showed that carefully selecting 9 Top Features among hundreds was enough to keep a performance 97\% as good as the original performance on the test set.
 
\begin{table*}
    \caption{Parameters of the 8 graph embedding methods.}
    \label{tab:parameters}
    \begin{tabular}{p{3.0cm} r r r r r r r r}
        \hline\noalign{\smallskip}
        \textbf{Parameter} & \textbf{FGSD} & \textbf{SF} & \textbf{G2V} & \textbf{DW} & \textbf{WL} & \textbf{N2V} & \textbf{BNE} & \textbf{GW} \\
        \noalign{\smallskip}\hline\noalign{\smallskip}
        \texttt{dimensions} & 200 & 128 & 128 & 128 & 32 & 128 & 8 & 100 \\
        \texttt{hist\_range} & 10 & - & - & - & - & - & - & -\\
        \texttt{wl\_iterations} & - & - & 1 & - & - & - & - & -\\
        \texttt{down\_sampling}  & - & - & $10^{-4}$ & - & - & - & - & -\\
        \texttt{learning\_rate} & - & - & 0.06 & 0.05 & 0.05 & - & - & -\\
        \texttt{epochs} & - & - & 12 & - & - & - & - & - \\
        \texttt{min\_count} & - & - & 1 & 1 & 1 & - & - & -\\
        \texttt{window\_size} & - & - & - & 10 & 4 & 10 & - & - \\
        \texttt{walk\_number} & - & - & - & 5 & 5 & 10 & - & -\\
        \texttt{walk\_length} & - & - & - & 80 & 80 & 20 & - & -\\
        \texttt{p} & - & - & - & - & - & 0.95 & - & -\\
        \texttt{q} & - & - & - & - & - & 1.0 & - & - \\
        \texttt{iterations} & - & - & - & - & - & - & 16 & - \\
        \texttt{order} & - & - & - & - & - & - & 1 & - \\
        \texttt{alpha} & - & - & - & - & - & - & 0.01 & - \\
        \texttt{step\_size} & - & - & - & - & - & - & - & 0.2 \\
        \texttt{heat\_coefficient} & - & - & - & - & - & - & - & 0.5\\
        \texttt{approximation} & - & - & - & - & - & - & - & 100\\
        \texttt{switch} & - & - & - & - & - & - & - & 1000 \\
        \noalign{\smallskip}\hline
    \end{tabular}
\end{table*}

The exact parameter settings used for all embedding methods are described in Table~\ref{tab:parameters}. Graph2vec is able to take into account a label associated to each node: we use the ID of the author modeled by the node. For BoostNE, it is worth stressing that parameter \texttt{order} has a strong effect on the performance, as increasing the value of this parameter lowers the performance. Using larger values might be beneficial on larger graphs, though. For GraphWave, performance stays relatively constant with a \textit{sample\_number} higher than 100, but strongly decreases with a lower value.

%%%%%%%%%%%%%%%%%%%%%%%%%%%%%%%%
\subsection{Performance on Abusive Message Detection}
\label{subsec:results}

%%%
\paragraph{Embeddings Only}
The first two columns of Table~\ref{tab:perfs} present the $F$-measure values obtained by our baseline and the 8 embedding methods described in Section~\ref{subsec:embeddings}. It also shows the dimension of the vector representations used to perform the classification. The last two columns correspond to results obtained when we use simultaneously the embedding methods and the topological measures from our baseline (described in Section~\ref{subsec:baseline}), by concatenating their vector representation. The reported performances are obtained following the protocol described in Section~\ref{subsec:DataProtocol}.

\begin{table*}
    \caption{$F$-measures obtained for the baseline and the 8 graph embedding methods. The left part corresponds to results obtained with the embedding methods alone, whereas the right part shows how they perform when combined with the topological measures used in the baseline. In the \textit{Scale} column, \textit{WG} stands for Whole-Graph and \textit{N} for Node.}
    \label{tab:perfs}
    \begin{tabular}{l p{1.2cm} r r@{\hspace{1.0cm}} r r}
        \hline\noalign{\smallskip}
        \textbf{Scale} & \textbf{Method} & \multicolumn{2}{r@{\hspace{1.0cm}}}{\textbf{Embeddings only}} & \multicolumn{2}{r}{\textbf{Embeddings \& Topo. meas.}} \\
        & & \textbf{Dimension} & \textbf{$F$-measure} & \textbf{Dimension} & \textbf{$F$-measure} \\
        \noalign{\smallskip}\hline\noalign{\smallskip}
        WG & FGSD & 200 & 77.06 & 659 & 87.27 \\
        & SF & 128 & 79.88 & 587 & 88.34 \\
        & Graph2vec & 128 & 81.91 & 587 & 89.16 \\
        \noalign{\smallskip}\hline\noalign{\smallskip}
        N & DeepWalk & 128 & 78.85 & 587 & 87.73 \\
        & Node2vec & 128 & 83.70 & 587 & 89.03 \\
        & Walklets & 128 & 79.49 & 587 & 88.15 \\
        & BoostNE & 136 & 63.28 & 595 & 86.54 \\
        & GraphWave & 200 & 83.04 & 659 & 87.97 \\
        \noalign{\smallskip}\hline\hline\noalign{\smallskip}
        -- & \textbf{Baseline} & Dimension: & \multicolumn{1}{l}{459} & $F$-measure: & \multicolumn{1}{l}{88.08} \\
        \noalign{\smallskip}\hline
    \end{tabular}
\end{table*}

We first focus on the results obtained without the topological measures (embeddings only). They demonstrate how appropriate the considered embedding methods are for the task at hand. Our first observation is that there is no clear distinction between node and whole-graph approaches in terms of $F$-measure. Since the whole graph is representing the message and its context, one could have thought that embedding the whole graph could allow capturing more important information than a single node embedding. However, it seems that these graphs can be well-characterized by focusing on the embedding representing only the node corresponding to the author of the \textit{targeted message}. This could mean that the relative position of this node in the graph is enough to characterize the whole conversation, and that node-level embeddings are able to capture such information.

The baseline yields the best performance, though. This is not surprising because this hand-crafted set of features was specifically designed for this task and dataset, whereas the embedding methods are somewhat generic. An interesting result is that Node2vec and GraphWave (node scale approaches), and Graph2vec (whole-graph scale) yield relatively good performance. These three approaches have several advantages over the baseline. First, they are not specifically designed for this task or dataset and are hence more likely to be efficient in other settings. Second, embedding methods are more scalable than hand-crafted sets of features. Computing the topological measures used in our baseline is computationally very expensive, with a total runtime of more than 8 hours. On the other hand, it only takes a few minutes to deal with Graph2vec, GraphWave and Node2vec on the same computer, which makes them a lot more time-efficient.
The other methods, except BoostNE, obtain correct performances with a $F$-measure around 79\%. SF and FGSD, which operate on the whole graph, might be penalized by the small size of our dataset and by the fact that graphs have approximately the same size and thus, possibly similar structures. DeepWalk is less efficient than Node2vec, which is in line with other studies~\cite{Goyal2018,Grover2016}. The Walklets algorithm learns multi-scale relationships in the graph. However, such relationships might not be very developed in our graphs, which could explain its lower performance. This observation could also be the reason of the very poor performances of BoostNE, which also operates on different granularity levels.

%%%%%%%%%
\paragraph{Embeddings \& Topological Measures}
Now, to study the complementarity between the baseline and the embedding methods, we propose to combine these features following the three fusion strategies proposed in~\cite{Cecillon2019} and summarized in Section~\ref{subsec:baseline}. An important difference with our previous work is that instead of using these strategies to combine topological measures and text features, our aim in this work is to combine topological measures and graph embeddings, since we do not use the textual content of the exchanged messages. 

Our experiments show that the three strategies lead to very similar performances on our task. Hence, in the remaining of this paper we present only the results obtained using one of them. We choose the \textit{Early fusion}, because it is the simpler of the three, and also because it eases interpreting the results, in particular regarding the feature ablation. In practice, the embedding generated by the considered embedding method is concatenated with the topological measures computed for the graph. We perform a new classification following exactly the same protocol as before, and report the obtained performance in the last two columns of Table~\ref{tab:perfs}.

Graph2vec, SF and Node2vec are the 3 methods that improve the baseline performance when combined with it, up to a 89.16 $F$-measure for Graph2vec. This result is critical, as it proves that the information captured by these three embedding methods is not similar, at least partially, to the information captured by the baseline. Furthermore, the additional information captured in the embeddings is useful for the classification task, since combining it with the baseline improves the classification performance. This result acknowledges the assumption that graph embedding methods can be used to detect abusive messages and that they can even capture useful information that is not caught by a hand-crafted set of measures.

When combined with the baseline, Walklets and Graphwave yield $F$-measures almost similar to what is obtained by the baseline alone. This seems to indicate that the generated embeddings contain information that is already captured by the baseline features, or that is useless for this specific classification task. However, even if the embeddings do not improve the performance, they do not introduce incorrect information (\textit{i.e.} noise) in the representations as the performances stays approximately the same with and without them.

Contrariwise, DeepWalk, FGSD and BoostNE combined with the baselines yield a $F$-measure that is inferior to that of the baseline on its own. It seems that the representations generated by these methods introduce some incorrect information when combined with the baseline, which causes a loss of performance. It is worth highlighting that these approaches were already the three worst performing methods when used without the baseline.

%%%%%%%%%%%%%%%%%%
\subsection{Feature Ablation}
\label{subsec:ablation}
In our previous work~\cite{Cecillon2019}, we identified, among the topological measures used in the baseline, the most discriminative features for our classification task, which we called \textit{Top Features}. For this purpose, we used a standard feature ablation process. 
%First, we train an estimator on the dataset to evaluate the importance of each feature in the classification. Then, we remove the least important feature from the set of features and train and test the classifier using only the remaining ones. That procedure is repeated on the reduced set of features, until only one is remaining. The latter a feature is removed, the more important this feature is in the classification process.
As embedding methods provide vector representations in which dimensions are not directly interpretable, it does not make sense to apply the same method here. Instead, we propose to study whether the most important topological measures from our baseline are well captured by the embedding methods.

\begin{figure*}[htb]
    \center
    \includegraphics[width=1\textwidth]{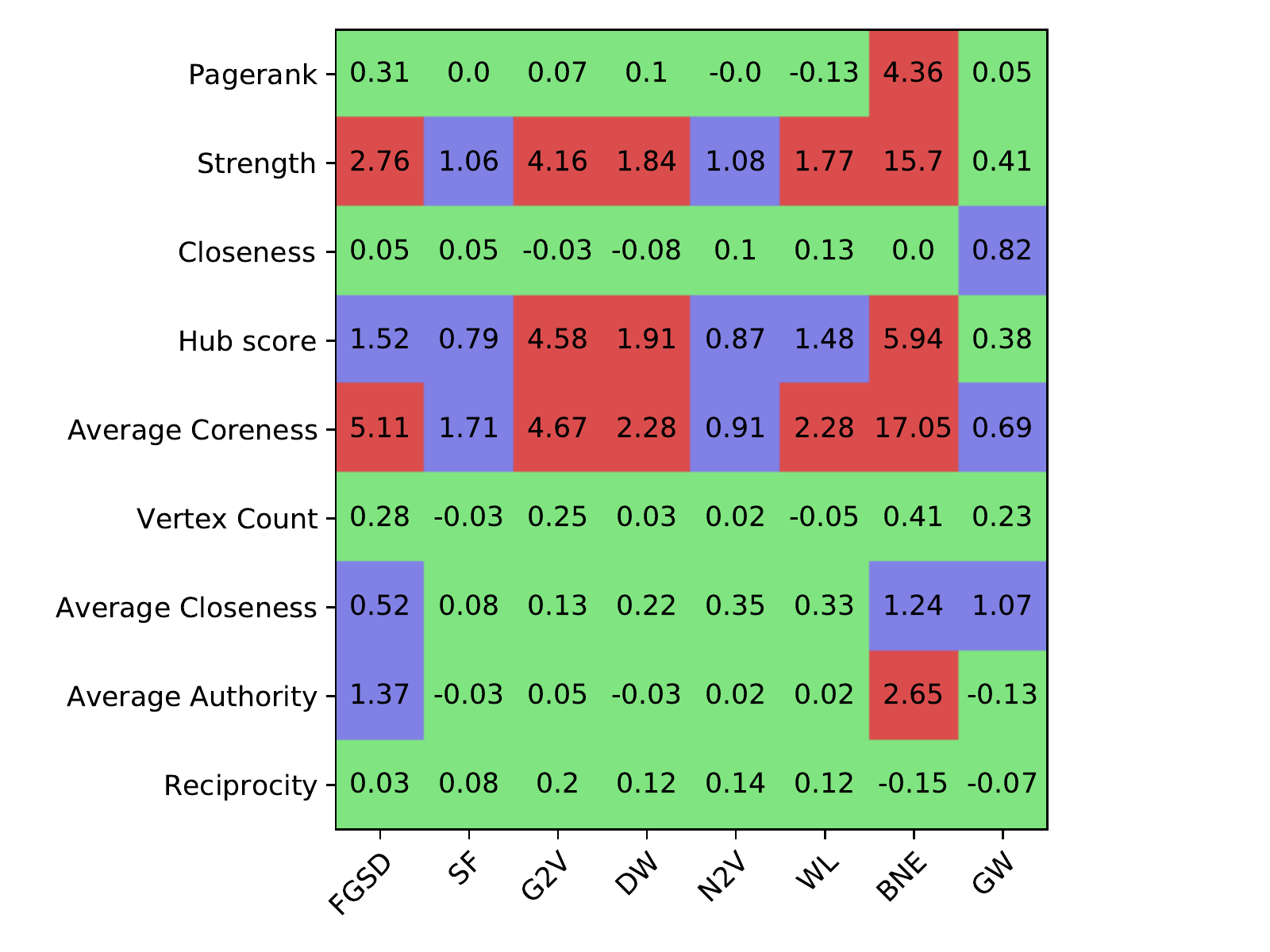}
    \caption{Topological measures captured (green), partially captured (blue) or not captured (red) by the embedding approaches. The first 4 topological measures are computed at the \textit{Graph} level and the last 5 topological measures are computed at the \textit{Node} level. Each value is the difference between the $F$-measure score obtained by the embedding method on its own and the score obtained by the embedding method completed by the corresponding Top Feature. Figure available at \href{https://doi.org/10.6084/m9.figshare.7442273}{10.6084/m9.figshare.7442273} under CC-BY license.}
    \label{fig:ablation}
\end{figure*}

To this end, we compare the $F$-measure score obtained by each embedding method on its own, with the score obtained by using a representation composed of the same embedding completed by one of the Top Features. Figure~\ref{fig:ablation} shows the difference between these two scores for all embedding methods and Top Features. If the performance significantly increases, we conclude that the topological measure was not captured by the embedding. This cases are represented in red in the figure. If the performance stays the same or increases by less than 0.50\% with the additional feature, we conclude that the structural property corresponding to this topological measure is well captured by the embedding (represented in green). If the performance increase is higher than 0.50\% but not statistically significant, we conclude that the property is only partially captured by the embedding method (represented in blue).

An interesting result shown by Figure~\ref{fig:ablation} is that some topological measures seem to be well captured by all, or almost all the embedding methods (\textit{e.g.} PageRank centrality, Vertex count, Closeness centrality at node and graph level and Reciprocity). Contrariwise, the average Coreness score, which is considered as a graph-scale measure, is partially captured by SF, a graph embedding method and by Node2vec and GraphWave, two node embeddings methods. The latter is also the only method that completely captures the Strength centrality and Hub score.

SF, Node2vec and GraphWave, which are among the best performing methods in Table~\ref{tab:perfs} are able to capture or partially capture all the important measures studied. However, they yield a $F$-measure much lower than the baseline. Therefore, we can suppose that these methods might not capture other properties of the graph which are less important but improve the performance when all combined. Graph2vec fails to capture some measures (\textit{i.e.} average Coreness score, Strength centrality and Hub score). However, as suggested by the result of its combination with the baseline, Graph2vec might miss some important measures of the baseline but this method is able to capture other important properties of the graph which are not conveyed by the baseline features. This is why Graph2vec is the best performing method when considering the combination with the baseline.  

An other interesting result of this study is that there is no clear difference in the type of information captured by node and whole-graph embedding approaches. Node embedding methods are able to capture certain graph-scale topological measures and whole-graph embedding methods can capture some node-scale measures. This property may result from the relatively small size of our graphs, as the second order neighborhood of a node might include a majority of the nodes in the graph. Thus, differences between node and whole-graph embedding methods are not as important as what they could be on larger graphs. Furthermore, our graphs are centered around a specific node. This specificity might help the whole graph embeddings to capture better node-level information.

%%%%%%%%%%%%%%%%%%%%%%%%%%%%%%%%%%%%%%%%%%%%%%%%%%%%%%%%%%%%%%%%
\section{Conclusion}
\label{sec:conclusion}
In this paper, we use graph embedding representations to tackle the problem of automatic abuse detection in online textual messages. We compare 8 methods operating on nodes and whole graphs to find the category of embeddings which fits the best the needs of this task. Our results show that Node2vec, GraphWave and Graph2vec are the methods that perform the best on this task. We compare the performance of these 8 graph embedding methods with a baseline previously designed using a feature engineering approach. With a 88.08 $F$-measure, this baseline outperforms the embedding methods, but the top ones obtain promising results: up to 83.70 with the Node2vec approach. We also study the complementarity between embedding methods and the topological measures used in the baseline. Combining them with \textit{Graph2vec} allows to improve the performance up to a 89.16 $F$-measure. Finally, we study which aspects of the graph structure each embedding method is able to capture. We find that methods operating on nodes and whole graphs are all able to include the information conveyed by certain topological measures defined both at node and graph scales.

A limitation of this work is the small size of our dataset (1,320 messages). Our application could benefit a larger dataset with more variety and examples of abusive messages. We have already started working in this direction, by proposing and freely distributing WAC\footnote{DOI: \href{https://doi.org/10.6084/m9.figshare.11299118}{10.6084/m9.figshare.11299118}}, a corpus based on Wikipedia edit discussion pages~\cite{cecillon2020wac}. It combines and improves two preexisting corpora to provide simultaneously comments annotated in terms of abuse and their surrounding conversation. This new corpus could be a larger field of experimentation, with around 383k annotated messages including 51k abusive ones distributed over 3 classes of abuse. Another limitation is the relatively small size of the graphs that we use to model conversations, which is likely to reduce the differences between node and whole-graph embedding methods. Working on larger graphs could help better distinguish the differences between these two types of embedding methods.

In the current work, we use static graphs to represent conversations. However, as our dataset contains details about the time at which messages were posted, a possible future work is to integrate a temporal aspect in our study. For example, by constructing sequences of embeddings to represent the evolution of conversation over time, or to experiment with representation able to simultaneously embed structural and temporal information. Another track is to leverage the content of messages through text embeddings, as we did previously with a feature engineering approach. Here too, it is possible to consider using separate embeddings for structure and text, or specific embeddings able to combine both types of information at once. Finally, another interesting track is to study the complementarity of different categories of graph embedding methods, for example by simultaneously using node, edge and whole graph representations.

%\begin{acknowledgements}
%\end{acknowledgements}

\section*{Conflict of interest}
On behalf of all authors, the corresponding author states that there is no conflict of interest.

\bibliographystyle{spmpsci}      % mathematics and physical sciences
\bibliography{Cecillon2020}   % name your BibTeX data base

\end{document}